# On the Main Factor That Causes the Instabilities of the Earth Rotation

Jin Sim[1], Kwan U Kim[1], Ryong Jin Jang[1], Jun-Sik Sin[2]

*Abstract:*

Earth rotation is one of astronomical phenomena without which it is impossible to think of human life. That is why the investigation on the Earth rotation is very important and it has a long history of study. Invention of quartz clocks in the 1930s and atomic time 1950s and introduction of modern technology into astronomic observation in recent years resulted in rapid development of the study in Earth's rotation.

The theory of the Earth rotation, however, has not been up to the high level of astronomic observation due to limitation of the time such as impossibility of quantitative calculation of moment of external force for Euler's dynamical equation based on Newtoniam mechanics. As a typical example, we can take the problems that cover the instabilities of the Earth's rotation proved completely by the astronomic observations as well as polar motion, the precession and nutation of the Earth rotation axis which have not been described in a single equation in a quantitative way from the unique law of the Earth rotation.

In particular, at present the problem of what the main factor causing the instabilities of the Earth rotation is has not been solved clearly in quantitative ways yet.

Therefore, this paper addresses a quantitative proof that the main factor which causes the instabilities of the Earth rotation is the moment of external force rather than variations in the relative atmospheric angular momentum and in moment of inertia of the Earth's body due to the time limitation and under some assumptions.

Then the future direction of research is proposed.

[1] the Central Information Agency for Science and Technology, *Democratic People's Republic of Korea;*

[2] Natural Science Center, **Kim Il Sung** University, *Democratic People's Republic of Korea;*;

Correspondence to: K.U.Kim, pyol_703@hotmail.com; J.-S. Sin, sinjunsik@163.com

## I. Simple Analysis of Euler–Liouville Equations and Statement of the Problems

Euler's dynamical equation

$$m_0 \frac{d^2\mathbf{r}}{dt^2} = \mathbf{F} = -\frac{Gm_0 M}{r^2}\mathbf{e}_r \quad \Rightarrow \quad \frac{d\mathbf{L}}{dt} = \mathbf{M} \tag{1}$$

that is well known in the history have been regarded as a cornerstones in research of the instabilities of the Earth rotation[1]-[5]. The equation is a different representation of Newton's Second Law [6]. Here **M** is the moment of external force out the Earth.

The equation can be written in various forms of letters.

$$\frac{dH}{dt} = M \tag{2}$$

Liouville processed Euler's dynamical equation and proposed Euler–Liouville equations [1], [6].

$$\frac{dH_i}{dt} + \varepsilon_{ijk}\omega_j H_k = M_i \tag{3}$$

That is, the following expressions can be written from (1)

$$\begin{aligned}
\frac{\dot{m}_1}{\sigma_r} + m_2 &= \varphi_2 \\
\frac{\dot{m}_2}{\sigma_r} - m_1 &= -\varphi_1 \\
\dot{m}_3 &= \dot{\varphi}_3
\end{aligned} \tag{4}$$

, while the following expressions from (2)

$$\begin{aligned}
\frac{1}{\sigma_r}\frac{dv_1}{dt} + v_2 &= \varphi_2 \\
\frac{1}{\sigma_r}\frac{dv_2}{dt} - v_1 &= -\varphi_1 \\
\frac{dv_3}{dt} &= \frac{d\varphi_3}{dt}
\end{aligned} \tag{5}$$

It is the different expression of the same thing. Explanation of the letters is the same as in [1-6]. Here, $\sigma_r$ and $\varphi_i$ can be obtained from the following expressions.

$$\sigma_r = \frac{C-A}{A}\Omega \tag{6}$$

$$\begin{aligned}\Omega^2(C-A)\varphi_1 &= \Omega^2 c_{13} + \Omega \dot{c}_{23} + \Omega h_1 + \dot{h}_2 - L_2 \\ \Omega^2(C-A)\varphi_2 &= \Omega^2 c_{23} - \Omega \dot{c}_{13} + \Omega h_2 - \dot{h}_1 + L_1 \\ \Omega^2 C \varphi_3 &= -\Omega^2 c_{33} - \Omega h_3 + \Omega \int_0^t L_3 dt\end{aligned} \tag{7}$$

, where dot "•" means $d/dt$. The left side of (4) is determined by the astronomical observations and the right side by the geophysical observations.

The equation of Earth rotation around its axis can be rewritten by integrating the third expression of (4)

$$m_z = \psi_z \tag{8}$$

, where $m_z = -\frac{\delta P}{P_0} = \frac{\delta \Omega}{\Omega_0}$, $P_0 = 86400 s$ and $\Omega = \Omega_0(1+m_z)$ is the Earth's instantaneous rotational angular velocity and $\psi_z$ is the excitation function.

The coordinate axis z is placed on the plane of meridian. The excitation function takes the following form from the third expression of (7)

$$\psi_z = \frac{1}{I\Omega_0^2}\left[-\Omega_0^2 \delta I - \Omega h_z + \Omega_0 \int_0^t M_z dt\right] \tag{9}$$

, where $M_z$ is the projection of the moment of external force on to z axis. I is averaged observations and $I + \delta I$ is the Earth's instantaneous moment of inertia with respect to z axis. $h_z$, projection of angular momentum rotating relatively with respect to z axis can be written as:

$$h_z = \int_V \rho\,(x\dot{y} - y\dot{x})\,dV = h_z(t) \tag{10}$$

*hz*, variation in angular momentum can also be written in the following form.

$$h_z = \int_V \varepsilon_{ijk} x_j u_k \rho\,dV \tag{11}$$

, where $\rho = \rho(x,y)$ is the density of substance at coordinates *x* and *y*.

As seen from Liouville's general solution (9), the factors which are probable to have an effect on the instabilities of the Earth rotation ($\psi_z$ known as the excitation function) are the variations in the atmospheric angular momentum, in the moment of inertia of the Earth's body and moment of external force. That is,

$$\psi_z = f[\,M_z,\,\Delta I_{33},\,\overline{\delta K_\alpha(i^d)} = h_z\,] \tag{12}$$

The previous researchers came up with the factors comparatively right, but they did not find the main factor which can influence the instabilities most greatly. They were mostly based on their abstract thought and reasoning and subjective points of view for analysis of the problem.

For example, Munk and Hassan studied the effect of variation in the moment of inertia ($\Delta I_z$) within the length of the day and they concluded that the degree of effect of variation in the moment of inertia on the instabilities of the Earth rotation is some percents of the degree of effect of variation in the atmospheric angular momentum.[3]. Therefore, the term was neglect-ted in (9), so after that the variation in the moment of inertia was neglected. Rapid progress in observational means led to discovery of the instabilities of the Earth rotation and short oscillation. Hence, many scholars took a great concern to the variation in moment of inertia.

While, the previous researchers removed the moment of external force in (9) by assuming that the Earth is an isolated system and neglected it by assuming that the moment of external force is much small if there is any. Furthermore, they have not found the numerical expressions which can represent the magnitude of the moment of external force in a quantitative way [4],[6],[7]. As a matter fact, Newtonian mechanics can give an exact explanation of the magnitude of the moment of external force in a physical way, but the former is impo ssible to find the latter correctly in a quantitative way and to give a mathematical expression of the latter since the latter is correctly expressed in terms of relativistic physical quantities which reflects the speed of light. Therefore, only relativistic dynamical system which involves Newtonian mechanics can solve the problem [8]-[10].

The previous researchers studied the Earth's rotation for a long time and have not completed it not because of their insufficient ability to solve it, but because of appearance of relativistic dynamics which can deal with the Earth's rotation in a quantitative way after 1905. Therefore, the scientific determination of the main factor among the moment of external force, the variation in the relative atmospheric angular momentum and the variation in the moment of inertia is very important for earth science. However, the previous researchers were not serious to treat the problem and have not completed it[11]-[13].

As a matter of fact, the Earth is not an isolated system.
It is a planet moving in the Solar system and is in spatial and rotational motion under the Sun's gravitational field. Therefore, as the spatial motion of the planets is determined by the Newton's law of universal gravitation in (1), so the rotational motion of the Earth is determined by the moment of external force for the Euler dynamical equation. Hence, the incompletion of the Earth's rotation problems is due to the time limitation in development of science and technology. Actually, at present, some problems remains unsolved as riddles for the instabilities of the Earth rotation because the former scholars have not solved the fundamental problem caused by the moment of rotation of external force which is mostly responsible for the Earth's rotation. The typical example is the irregular and "peak" change in the Earth's rotation velocity the reason for which has not been accounted for until now[14],[15].

The irregular and "peak" change in the Earth's rotation velocity can make the length of the day to change up to 0.0034s.
The greatest change happened in 1864, 1878, 1898 and 1920 since 1820[14]. It is determined by observation of the Moon and as a riddle, the cause for the astronomical phenomena has not been explained yet. Newcomb said that it is the strangest astronomical phenomenon[2]. The causes for secular deceleration of the Earth's rotation (secular change) and periodic change in the Earth's rotation velocity have not been accounted for yet. At present, a great number of researchers believed that the secular deceleration was caused by oceanic tides and the periodic change by atmospheric periodic change.

Another typical example can be taken for secular motion of the mean pole on the Earth(change per century)[17]. In other words, the mean pole shifts at a velocity of about 10 cm/year on the Earth. In spite of existence of internation al research organizations for study in the secular change of the mean pole on the Earth, no theory has accounted for the reason for the astronomical phenomena [16],[17].

Furthermore, the International Earth Rotation and Reference Systems Service (IERS) does not know well about the Law of the Earth Rotation Change which can explain not only the instabilities of the Earth's rotation velocity, but also precession, nutation of the Earth's rotation axis and polar motion in the body of the Earth in a quantitative way from the single dynamical equation.

The previous theory of the Earth rotation constructed based on Newtonian dynamics cannot account for the observed astronomical phenomena for the Earth rotation as have been already mentioned above at all.

Then, why have the results from the previous theory of the Earth rotation not coincided with the observational data obtained by modern observational means?

It is because the theory of the Earth's rotation has been confined in the range of Euler's theory of the Earth's rotation constructed based on Newtonian mechanics which is defective in explaining and studying the relativistic phenomena of gravitation. In other words, Newtonian dynamics cannot account for the perihelion advance in the planetary orbits and the same goes for the reasons for the instabilities of the Earth rotation.

The authors made a profound study in them in the frame of relativistic dynamical system involving Newtonian dynamics based on the right physical laws in accordance with the physical methodology and discovered the law for the Earth's real rotation (the Law of the Earth Rotation Change).

Therefore, the paper addresses the main factor causing the instabilities of the Earth rotation in a quantitative way using the Law of the Earth Rotation Change found in relativistic dynamical system. In addition to it, it presents the application of the Law of the Earth Rotation Change in the future.

## II. The Factors Obtained from the Law of the Earth Rotation Change Found in Relativistic Dynamical System.

As a matter of fact, the literature review of the previous studies in theory of the Earth's rotation shows that none of them has gone beyond the range of the system of the equations formed in Newtonian dynamical system in order to investigate the Earth rotation.

However, the theory formed in Newtonian dynamical system is impossible to treat the complicated observational problems related with the instabilities of the Earth's rotation velocity in a single dynamical system which have been found and confirmed by rapid development of the observational equipment. In other words, theory of the Earth's rotation has not kept up with the high-accuracy astronomical observations. It is the greatest shortcoming of the theory which is due to the previous researchers not having done studied the theory of the Earth's rotation in a quantitative way in system of relativistic theory of gravitation in spite of long period of over 100 years since construction of relativistic dynamics involving Newtonian mechanics and general theory of relativity. The previous researchers have not put their attention to theory of the Earth's rotation based on relativistic theory of gravitation at once after it appeared. Therefore, the authors have studied theory of the Earth's rotation in a quantitative way based on the fundamental law of relativistic dynamics involving Newtonian mechanics in it [8], [9].

It is well known that the Earth is in constant motion under gravitational interaction with the planets such as the Sun and Moon in the Sun's gravitational field and the cosmic objects.

In other words, the Earth is not an isolated system. Therefore, the celestial bodies, such as the Sun and Moon have an effect on the Earth's motion in the space, while moment by external force is certain to act on the Earth's body to have an effect on the Earth's rotation. However, the former researchers have not found the expression which is possible to present the magnitude of the moment by external force in .a quantitative way yet.

It is the greatest shortcoming of the previous theory on the Earth's rotation.

The Earth moves round the Sun along the real elliptic orbit the perihelion o f which advances according to the real law of relativistic motion

$$\frac{d}{dt}\left[\frac{m_0 \frac{d\mathbf{r}}{dt}}{\sqrt{1-\frac{1}{c^2}\left(\frac{d\mathbf{r}}{dt}\right)^2}}\right] = \frac{d}{dt}\left[m\frac{d\mathbf{r}}{dt}\right] = \frac{d\mathbf{P}}{dt} = \mathbf{F} = -\frac{Gm_0 M}{r^2}\mathbf{e}_r - \frac{5G^2 m_0^2 M^2}{\varepsilon d^3}\mathbf{e}_r + \cdots + \sum_j \mathbf{F}_j^{pert}(\mathbf{r}_i) \tag{13}$$

in the Sun's central force field added up by perturbative force, while the Earth rotates on its axis[8]. That is, the Earth rotates due to the moment by external force produced by gravitational interaction between the celestial bodies such as the Sun and Moon.

Therefore, application of the law of relativistic motion (13) to theory of the Earth's rotation provides the theoretical conclusion that the moment by external force produced by the celestial bodies out the Earth acts on the Earth and has a key effect on the Earth's rotation. That is, the Law of the Earth

Rotation Change is clearly derived from the Earth's law of relativistic motion [8]

$$\frac{d\mathbf{L}}{dt} = \mathbf{M} = -\frac{[1+\frac{5\alpha}{Er_g}]}{\sqrt{1+\left(\frac{1}{r_g}\cdot\frac{dr_g}{d\varphi}\right)^2}} \cdot K(X) \cdot \frac{5\alpha^2}{Er_g^2} \cdot \mathbf{e}'_z + \mathbf{M}^{pert}_{Moon} + \mathbf{M}^{pert}_{plan}(i) + \cdots \quad (14)$$

The right side in the expression is the mathematical representation of the moment by external force acting on the Earth.

As seen from (1), the physical idea was set up by Euler based on the Newton's Second Law.

However, as seen from (14), the magnitude of the key moment ( м ) by external force produced by the Sun cannot be obtained in a quantitative way from Newtonian mechanics since it can be presented in a quantitative way only by means of the physical quantities of relativistic dynamics.

The previous researchers' having not presented the magnitude of the moment by external force acting on the Earth in a quantitative way is due to the time limitation in the previous theory before relativistic theory of gravitation was constructed. The authors found the magnitude of the key moment by external force produced by the Sun in the quantitative way on the basis of relativistic dynamics and the Law of the Earth's relativistic motion. It is the difference of the author's findings from the previous researcher's ones.

One can see from the Law of the Earth Rotation Change that the Earth rotates affected by the key moment by the Sun's gravitational attraction and by the perturbation moment produced by the Moon and the planets.

The Earth rotates with it's rotation axis inclined to the ecliptic at an angle of $23°\,27'$ and the former is in the precession and nutation with respect to the latter.

Therefore, the Law of the Earth Rotation Change can be solved into two methods.

One is to divide the external torque in the rotary component and the vertical component perpendicular to the axis of the Earth's rotation by projecting the former directly onto the latter. Then, you can find the rotational component of the Earth, that is, the equation of the Earth's rotation as:

$$\frac{d}{dt}[I_0\omega_0] = M = -\frac{5\alpha^2 a^2}{\varepsilon_0 b^4} f_M(t) + \sum_j M^{pert}_{j\omega} \quad (15)$$

At this time, all the astronomical phenomena which are likely to occur in the Earth's rotation should be obtained, the theoretical values of them calculated and the theoretical values and astronomical observations compared and interpreted for the experimental verification of the theory. The other one is to connect the dynamical equation and kinematic one followed from projection of the Law of the Earth Rotation Change to the coordinate system of motion and to solve the derived system of equations

$$A\ddot{\psi}\sin^2\theta + 2A\dot{\psi}\dot{\theta}\sin\theta\cos\theta - C\omega_z\dot{\theta}\sin\theta = M_{Sun}\sin^2\theta + M_{Moon}^{pert}\sin\theta\sin\theta_i\sin(\varphi_i - \varphi) + \cdots \quad (16)$$

$$A\ddot{\theta} - A\dot{\psi}^2\sin\theta\cos\theta + C\omega_z\dot{\psi}\sin\theta = 0 \qquad + M_{Moon}^{pert}\sin\theta_i\sin(\varphi_i - \varphi) + \cdots$$

$$C\frac{d\omega_z}{dt} = M_{Sun}\cos\theta + M_{Moon}^{pert}\cos\theta_i \qquad + \cdots$$

Figure 1    **Earth's Rotation Verified by Astronomical Observations**

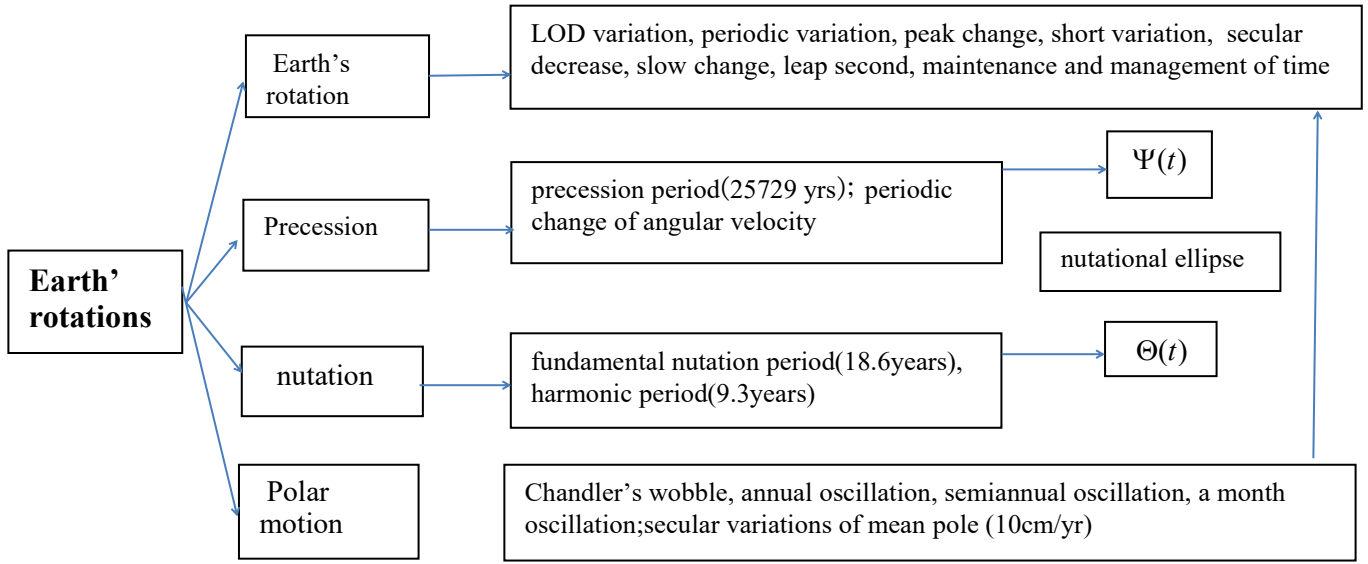

, that is, to combine the equation of precession, nutation and rotation and to solve it [9]. Here, all the astronomical phenomena likely to occur not only in the Earth's rotation but also in precession, nutation and polar motion should be given in a theoretical way, the theoretical values of them calculated and the theoretical values and astronomical observations compared and analysed for the experimental verification of the theory.

Speaking the conclusion beforehand, not only the instabilities of the Earth's rotation rate, but also precession, nutation of the Earth's rotation axis and polar motion can be treated through the solution of the single dynamical equation (14) in a quantitative way and all of the causes for the astronomical phenomena can be given and accounted for by the factors in the equation. The Earth's rotation verified up to now by the astronomical observations is shown in Figure 1.

However, all the reasons for the astronomical phenomena have not been successfully explained yet.

As a matter of fact, the "Journées 2019 - Astrometry, Earth Rotation and Reference Systems in the Gaia era" held in France did not give the correct answers to the causes for the astronomical phenomena. After all, theory of the Earth's rotation originated from Newtonian mechanics has not matched the high- accuracy astronomical observation yet.

However, the Law of the Earth Rotation Change found from the law of the Earth's motion for relativistic dynamics can account for all the astronomical phenomena known by the observation of the Earth's rotation through the single Law of the Earth's Rotation Change is shown in Figure 1[8], [9].

The fact shows an objective verification that the authors have discovered the real law of the Earth's rotation based on the physical laws of relativistic dynamics.

In addition, it shows in a quantitative way that the main factor causing the instabilities of the Earth's rotation is the moment by external force which the previous researchers have ignored.

By solving the Law (14) of the Earth Rotation Change around its axis follo wed from the law of the Earth's rotation (15) and calculating the relative varia tion of the physical quantities, the following general solution is obtained [8].

$$\frac{\overline{\Delta \omega_{0E}(i^d)}}{\omega_{00E}} = -\frac{A_E}{T_{00E}} \cdot \frac{1}{N} \cdot \int_{(i-1)N^s}^{iN^s} f_M(t)\, dt + \frac{\Delta T_{00E}}{T_{00E}} - \frac{1}{N}\sum_{i=1}^{N} \frac{\Delta I_0(i)}{I_0(i)} + \frac{T_{00E}}{2\pi I_{00}} \cdot \frac{1}{N} \cdot \int_{(i-1)N^s}^{iN^s} \sum_j M_\varphi^{\text{pert}}\, dt - \frac{\overline{\delta K_\alpha(i^d)}}{I_{00}\, \omega_{00E}}$$

$$10^{-8} \qquad 10^{-8} \qquad \sim 10^{-8} \qquad 10^{-12} \qquad 10^{-13} \qquad 10^{-14}$$

$$= -\frac{\overline{\Delta T_{0E}(i^d)}}{T_{00E}} \equiv m_z = \psi_z = \frac{1}{\Omega_0 I} \int_0^t M_z\, dt - \frac{\overline{\Delta I_{33}}}{I} - \frac{h_z}{\Omega_0 I}$$

$$10^{-8} \qquad 10^{-8}, 10^{-8}, \quad 10^{-8} \qquad 10^{-13} \qquad 10^{-14} \qquad (17)$$

, where the left side of $m_z$ is the result of the solution to our equation and the right side of it is the result of Liouville equation. Below the numerical expression, the range of the physical quantities to be analyzed are given. The terms in (17) correspond to each other. And variation in moment of inertia is $\delta I_0'(i^d) = \overline{\Delta I_{33}}$ and variation in the angular momentum of the atmosphere is $\overline{\delta K_\alpha(i^d)} = h_z$.

The rest terms reflect the portion of contribution of the moment of external force to the excitation function $\psi_z$. It should be kept in mind in (17) that the observation $m_z$ is *LOD* of *UT*1 caused by the instabilities of the Earth rotation, that is, the relative value of $D$ divided by the atomic time ($86400\,s$) and the value is in the following range

$$\frac{\overline{\Delta\omega_0(i^d)}}{\omega_{00E}} = -\frac{\overline{\Delta T_{0E}(i^d)}}{T_{00E}} = \frac{LOD}{86400\,s} = m_z = 10^{-8} \sim 10^{-9} \tag{18}$$

from the stability of the Earth's rotation. Therefore $m_z$ is the dimensionless argument. Hence, the goal of the paper is to find the main factor that causes the instabilities of the Earth's rotation by analyzing the factors in (17) in a quantitative way.

### III. Effect of variations in the Earth's atmospheric angular momentum on the Earth rotation

Seasonal variations in the velocity of the Earth rotation were discovered by comparison of the report of a high-precision quartz clock with the astronomic observations [6], [18].
A lot of suppositions and assumptions were proposed to explain the astronomic phenomenon.
The suppositions and assumptions that the atmosphere of the Earth has a main effect on the seasonal variations in the Earth rotation among the factors have been mentioned in not a few publications [3]-[6], [19]-[22].
Is the viewpoint right, indeed?

Two methods are possible for the theory [1]-[5], [22],[23], the first of which describes the balance of angular momentum of the rotating atmosphere (angular momentum method) and second one accounts for the mechanical effect of the atmosphere on the Earth's crust (method of the moments of forces).

### A. Angular momentum method (Balance method)

The first method is based on the assumption that the geosphere–atmosphere is the isolated system, namely,

$$I\Omega + K_\alpha = const \tag{19}$$

, where $K_\alpha$ is the absolute angular momentum of the atmosphere and $I\Omega$ is the absolute angular momentum of the Earth without the atmosphere.
From (19), the following expression is obtained.

$$\frac{\delta \Omega}{\Omega_0} = -\frac{1}{I\Omega}\delta K_\alpha \tag{20}$$

The assumption that the geosphere–atmosphere is the isolated system and the fact that the variation in the moment of inertia $\delta I$ is very small make us to describe the excitation function
(9) in a simplified form as follows.

$$\psi_z = -\frac{1}{\Omega_0 I}\int_V \rho(x\dot{y}-y\dot{x})\,dV = -\frac{h_z}{\Omega_0 I} \tag{21}$$

Munk and Hassan studied the effect of $\Delta I_z$ within the length of day in 1961 and they concluded that it was a several percent of the angular momentum at most [3].
Therefore, we can see that the basic expression of $\delta K_\alpha$, variation in the absolute angular momentum of the atmosphere can be determined according to the relative angular momentum $h_z$ from the equality of the left side of (20) and (21) [2].
That is,

$$\overline{\delta K_\alpha (i^d)} = h_z = \int_V \rho(x\dot{y}-y\dot{x})\,dV \tag{22}$$

Integration is carried out over the volume of the atmosphere $V$.
The excitation function takes the following form in the spherical coordinates.

$$\psi_z = -\frac{1}{\Omega_0 I}\int_V \rho r^3 Cos^2\theta \cdot V_\varphi \, dt\, d\theta \tag{23}$$

It can be described as follows [3].

$$\psi_z = -\frac{2\pi}{\Omega_0 I}\int_{R_E}^{R_E+H}\int_{\theta=-\frac{\pi}{2}}^{\theta=\frac{\pi}{2}} \rho(r)\, r^3 Cos^2\theta \cdot \overline{V_\varphi(r,\theta)}\, dt\, d\theta \tag{24}$$

Here, $H$ is the upper limit of the integral with respect to the height of the atmosphere.

It is said that the main phenomenon contributing to the total excitation function is the circulation of the global zonal winds in the atmosphere up to a height of 30km above sea level [3].

### B. Method of the moments of forces

The method of the moments of forces is based on the following equation.

$$\frac{d}{dt}[I\Omega + K_\alpha] = x(t) \neq 0 \qquad (25)$$

That is,

$$\frac{\delta\Omega}{\Omega_0} = -\frac{1}{I\Omega_0}\delta K_\alpha + \frac{1}{I\Omega_0}\int_{t_1}^{t_2} x(\tau)d\tau \qquad (26)$$

, where $x(\tau)$ is a function corresponding to the process acting on the atmospheric circulation from the outside.

And $\delta\Omega = \Omega - \Omega_0$, $\delta K_\alpha = K_\alpha - K_{\alpha 0}$, $\Omega$ and $K_{\alpha 0}$,

integral constants are the averaged values.

To calculate the effect of the atmospheric circulation on the velocity of the Earth rotation by the method of the moments of forces, 6 assumptions were made and the variation in the angular velocity of the Earth rotation due to the mechanical interaction between the atmosphere [24] and the Earth can be described in the following equation.

$$\frac{d\omega}{dt} = -\frac{1}{I}\int_0^\pi \int_0^{2\pi} \tau\, r^3 Sin^2\theta\; d\lambda\, d\theta \qquad (27)$$

The seasonal variations in the Earth rotation have not been solved completely yet, because the problems of the mechanical action of the atmosphere on the Earth's surface cannot be formulated practically and correctly and the angular momentum method is based on the assumption that geosphere–atmosphere is the isolated system [2],[22].

## C. Quantitative analysis of relative angular momentum of the atmosphere $h_z$

Some scholars mentioned as if the seasonal oscillations could be explained by the relative angular momentum $h_z$ [4], [5], that is,

$$h_3 = \left[144 + 23 Cos(2\oplus - 20°) + 14 Cos(2\oplus - 222°)\right] \times 10^{24} Kg m^2 s^{-1} \tag{28}$$

Can it be the right physical quantity as called the relative angular momentum of the atmosphere, indeed?

In addition, the author can directly obtain the following result from (17) in accordance with the assumption that the Earth is an isolated system in (17) and the conclusion of little change in $\Delta I_{33}$ by Munk and Hassan.

$$-\frac{\overline{\delta K_\alpha (i^d)}}{I_{00} \omega_{00E}} = -\frac{\overline{\Delta T_{0E}(i^d)}}{T_{00E}} = m_z = \psi_z = -\frac{h_z}{\Omega_0 I} \tag{29}$$

Now, can we say that the angular momentum method (21) is, indeed, based on the right physical vision?

In order to solve the problem, let us make an analysis of the physical meaning of $h_z$ as follows

$$h_z = \delta K_\alpha = K_\alpha - K_{\alpha 0} = \int_V (r \times u)\rho \, dV = \int_V \rho(x\dot{y} - y\dot{x})\, dV = \int_V \varepsilon_{ijk} x_j u_k \rho \, dV \tag{30}$$

Here, the relative angular momentum of atmosphere is the ratio as

$$\frac{\delta K_\alpha}{K_{\alpha 0}} = \frac{K_\alpha - K_{\alpha 0}}{K_{\alpha 0}} = \frac{h_z}{K_{\alpha 0}} = \frac{1}{K_{\alpha 0}} \int_V (r \times u) \rho \, dV = \frac{1}{K_{\alpha 0}} \int_V \rho(x\dot{y} - y\dot{x}) dV = \frac{1}{K_{\alpha 0}} \int_V \varepsilon_{ijk} x_j u_k \rho \, dV \tag{31}$$

from the physical point of view. That is, the correct meaning of the ratio is precisely the relative angular momentum of atmosphere. The physical quantity is dimensionless.

However, $h_z$ in (30) has the dimension of angular momentum. Therefore, wrong name was made for $h_z$. As you can see in (30), $h_z$ is the variation in the atmospheric angular momentum. The maximum of the total angular momentum which the atmosphere can have will not exceed $1.5 \times 10^{35} gCm^2 s^{-1}$.

That is, the value is calculated from the following expression, atmospheric angular momentum:

$$K_\alpha = \sum_i m_i R_i^2 \omega_{0i} \leq K_\alpha^{max} = I_W \omega_z = m_W R_E^2 \omega_z = 5.1 \times 10^{21} \times (6.371 \times 10^8)^2 \times \frac{2\pi}{86400} = \quad (32)$$

$$= 1.5053 \times 10^{35} [gCm^2 s^{-1}] = 1.5 \times 10^{28} [Kg\ m^2 s^{-1}]$$

The value is very small in comparison to the Earth's angular momentum. That is,

$$\frac{K_\alpha^{max}}{I\Omega_0} = \frac{1.5053 \times 10^{35}}{8.068 \times 10^{44} \times \frac{2\pi}{86400}} = \frac{1.5053 \times 10^{35}}{5.8672 \times 10^{40}} = 2.5656 \times 10^{-6} \ll 1 \quad (33)$$

, where the change of the atmospheric absolute angular momentum $K_\alpha$ with respect to time can be calculated as follows.

$$\frac{dK_\alpha}{dt} = \frac{d}{dt}[I_W \omega_Z] \leq I_W \cdot \frac{d\omega_Z}{dt} = \frac{dK_\alpha^{max}}{dt} = 2.070071691 \times 10^{39} \times 3.26 \times 10^{-12} = \quad (34)$$

$$= 6.748433713 \times 10^{27} [erg] = 6.74843 \times 10^{20} J$$

The maximum resulted from the use of the following maximal moment of inertia which the atmosphere can have on the body of the Earth as:

$$I_W = m_W \times R_E^2 = 5.1 \times 10^{21} g \times (6.371 \times 10^8 Cm)^2 = 2.070071691 \times 10^{39} [gCm^2]$$

As will be seen later, $\frac{d\omega_z}{dt}$ is the angular acceleration of the Earth's rotation (See (60)). The authors can clearly state from (34) that although the atmospheric angular momentum is possible to change at such rate as

$$\frac{dK_\alpha}{dt} \leq \frac{dK_\alpha^{max}}{dt} = 6.748433713 \times 10^{27} [erg] \quad (35)$$

at each instant in time, it cannot be greater than the maximum. As a result, we can conclude that the magnitude of the variation $h_z$ in the atmospheric angular momentum will not exceed the theoretical value as:

$$h_z = dK_\alpha \leq dK_\alpha^{max} = 6.748433713 \times 10^{27} [gCm^2 s^{-1}](\frac{dt}{1s})] = 6.74843 \times 10^{27} [gCm^2 s^{-1}] \quad (36)$$

(where $dt = 1s$)

In view of the expressions mentioned above, you can see that the value of the relative variation (31) of the atmospheric angular momentum can be written as:

$$-\frac{dK_\alpha^{max}}{K_\alpha^{max}} = \frac{6.74843 \times 10^{27} [gCm^2 s^{-1}]}{1.5053 \times 10^{35} [gCm^2 s^{-1}]} = 4.48312 \times 10^{-8} \ll 1 \quad (37)$$

And the ratio of the atmospheric angular momentum of the Earth to the angular momentum of the body of the Earth can be written as:

$$-\frac{K_\alpha^{max}}{\Omega_0 I} = \frac{1.5053 \times 10^{35} [gCm^2 s^{-1}]}{5.8672 \times 10^{40} [gCm^2 s^{-1}]} = 2.56562 \times 10^{-6} \ll 1 \quad (38)$$

Through the (37) and (38), you can see that there exists the following clear quantitative relationship between these physical quantities:

$$dK_\alpha^{max} \ll K_\alpha^{max} \quad , \quad K_\alpha^{max} \ll \Omega_0 I \tag{39}$$

The authors can give the quantitative answer to (29) mentioned above using the objective numerical values.

The left side of $m_z$ in (29) is the observation of the Earth rotation.

You know from the stabilities of the Earth's rotation that the observation is in the range of

$$-\frac{\overline{\Delta T_{0E}(i^d)}}{T_{00E}} = \frac{LOD}{86400s} = m_z = 10^{-8} \sim 10^{-9} \tag{40}$$

Therefore, the result in the right side of the excitation function $\psi_z$ in (29)

$$\psi_z = -\frac{h_z}{\Omega_0 I} \tag{41}$$

is the theoretical value of the excitation function and it is necessary for explanation of the astronomical phenomena of $m_z$.
Only when the result is equal to the observation of $m_z$ you can say that the result is the right one.

Therefore, let us estimate the magnitude of the right side of (41) in terms of (36),(37) and (38).

That is, the following result is obtained.

$$\psi_z = -\frac{h_z}{\Omega_0 I} = \frac{-dK_\alpha}{\Omega_0 I} \leq \frac{-K_\alpha^{max}}{\Omega_0 I} \times \frac{dK_\alpha^{max}}{K_\alpha^{max}} = 2.56562 \times 10^{-6} \times 4.48312 \times 10^{-8} = \tag{42}$$

$$= 1.150196638 \times 10^{-13} \neq 10^{-8} \sim 10^{-9} = -\frac{\overline{\Delta T_{0E}(i^d)}}{T_{00E}} = \frac{D}{86400s} = m_z$$

As you can see, first above all, the numerical value is not equal to the observation and the former is too much different from the latter.

If you use $10^{-19} \sim 10^{-20} [rad/s^2]$, angular acceleration proposed by Sidorenkov, you can obtain the following result.

$$\frac{dK_\alpha}{dt} = \frac{d}{dt}[I_W \omega_Z] \leq I_W \frac{d\omega_z}{dt} = \frac{dK_\alpha^{max}}{dt} = 2.07 \times 10^{39} \times 10^{-19} =$$

$$= 2.07 \times 10^{20} [erg] = 2.07 \times 10^{13} J$$

$h_z = dK_\alpha$ will be equal to $2.07 \times 10^{20} gCm^2 s^{-1}$ [4].

Therefore, the variation in the relative angular momentum of the atmosphere can be written as:

$$-\frac{dK_\alpha^{max}}{K_\alpha^{max}} = \frac{2.07 \times 10^{20}[gCm^2 s^{-1}]}{1.5053 \times 10^{35}[gCm^2 s^{-1}]} = 1.38 \times 10^{-15} \ll 1$$

The relative ratio is very small compared to (37).
As a result, $\psi_z$, excitation function, can be written as:

$$\psi_z = -\frac{h_z}{\Omega_0 I} = \frac{-dK_\alpha}{\Omega_0 I} \leq \frac{-K_\alpha^{max}}{\Omega_0 I} \times \frac{dK_\alpha^{max}}{K_\alpha^{max}} = 2.56562 \times 10^{-6} \times 1.38 \times 10^{-15} =$$

$$= 3.5 \times 10^{-21} \underset{\neq}{\ll} 10^{-8} \sim 10^{-9} = -\frac{\overline{\Delta T_{0E}(i^d)}}{T_{00E}} = \frac{D}{86400 s} = m_z$$

, which is much different from the value in (42)
$(3.5 \times 10^{-21} \neq 10^{-8} \sim 10^{-9})$

The quantitative relation demonstrates that the instabilities of the Earth's rotation ($m_z$) cannot be explained in terms of the variations in the atmospheric angular momentum. It manifests that $h_z$, variation in the atmospheric angular momentum is not the main variable determining the excitation function $\psi_z$ in (17). In addition, it follows from (42), that is,

$$-\frac{h_z}{\Omega_0 I} = \frac{-dK_\alpha}{\Omega_0 I} \leq \frac{K_\alpha^{max}}{\Omega_0 I} \times \frac{-dK_\alpha^{max}}{K_\alpha^{max}} = 1.150196638 \times 10^{-13} \quad (43)$$

that the maximal variation value which $h_z$ can take should not be greater than the following expression:

$$-h_z = -\delta K_\alpha \leq -dK_\alpha^{max} = \Omega_0 I \times 1.15096638 \times 10^{-13} =$$
$$= 5.8672 \times 10^{40} \times 1.15096638 \times 10^{-13} = \quad (44)$$
$$= 6.74843 \times 10^{27}[gCm^2 s^{-1}] = 6.74843 \times 10^{20}[Kg\, m^2 s^{-1}]$$

However, the theoretical value of $h_z$ is exceedingly great in [4], [5]. For example, the values of $h_z$ in (28) are different as much as 200,000 times in comparison to the values in (44).

$$\begin{aligned}\delta K_\alpha &= h_z(\text{contant amplitude}) &&= 144 \times 10^{24}[Kg\, m^2 s^{-1}] = 1.44 \times 10^{33}[gCm^2 s^{-1}]\\ \delta K_\alpha &= h_z(\text{annual amplitude}) &&= 23 \times 10^{24}[Kg\, m^2 s^{-1}] = 2.3 \times 10^{32}[gCm^2 s^{-1}]\\ \delta K_\alpha &= h_z(\text{semi-annual amplitude}) &= 14 \times 10^{24}[Kg\, m^2 s^{-1}] = 1.4 \times 10^{32}[gCm^2 s^{-1}]\end{aligned} \quad (45)$$

The expression (35), physical analysis of values of $h_z = \delta K_\alpha$ shows that the value of $h_z$ cannot be greater than $6.74843 \times 10^{27}[gCm^2 s^{-1}]$.

However, the values of (45) is 200,000 times greater than the limit. You can find a typical examples on the page 128 and 131 in [4], which leads to a theoretical contradiction.

In addition, when using $10^{-19} \sim 10^{-20}[rad/s^2]$, the angular acceleration of the Earth's rotation proposed by Sidorenkov, the authors obtained the variation in the atmospheric angular momentum, that is, $h_z = dK_\alpha = 2.07 \times 10^{20} gCm^2 s^{-1}$, which should be equal to the value in (45) in principle.

But, as you can see, it is much different from it ($10^{12} times$). It shows that the previous theory is in self-contradiction.

Then, you can ask from where and how (28), the value of (45) is derived.

Dividing the theoretical values of $h_z$ by the angular momentum of the body of the Earth gives the following relative ratio of $m_z$:

$$\frac{h_z(\text{constant amplitude})}{\Omega_0 I} = \frac{1.44 \times 10^{33}}{5.8672 \times 10^{40}} = 2.46 \times 10^{-8}, \quad \sim m_z \approx 10^{-8}$$

$$\frac{h_z(\text{annual amplitude})}{\Omega_0 I} = \frac{2.3 \times 10^{32}}{5.8672 \times 10^{40}} = 3.92 \times 10^{-9}, \quad \sim m_z \approx 10^{-9}$$

$$\frac{h_z(\text{semi-annual amplitude})}{\Omega_0 I} = \frac{1.4 \times 10^{32}}{5.8672 \times 10^{40}} = 2.39 \times 10^{-9}, \quad \sim m_z \approx 10^{-9}$$

It shows that the theoretical values of $h_z$ in [4],[5] has intentionally been fabricated to explain

$$-\frac{\overline{\Delta T_{0E}(i^d)}}{T_{00E}} = \frac{D}{86400 s} = m_z = 10^{-8} \sim 10^{-9},$$

the instabilities of the Earth rotation. Furthermore, in view of the fact that the absolute angular momentum of the atmosphere does not exceed the value in (32)

$$K_\alpha \leq K_\alpha^{max} = 1.5053 \times 10^{35}[gCm^2 s^{-1}] = 1.5 \times 10^{28}[Kg\, m^2 s^{-1}]$$

, it is possible to say that the theoretical value in (45), the variation of $hz$ is too much great.

If the atmosphere has an effect on the Earth rotation, its physical mechanism is the exchange in the angular momentum and the great variation in the angular momentum in (36) would remove the atmospheric angular momentum of the Earth in about ten minutes, that is,

$$\left(1.5053 \times 10^{35}[gCm^2 s^{-1}] \div 2.3 \times 10^{33}[gCm^2 s^{-1}]/s = 652\, s \cong 10.87\, min\right)$$

, which is the contradiction to natural phenomena and means that such great change as in (45) is impossible to occur.

In addition, it is the contradiction to the statement "The relative angular momentum $h_z$ are also assumed to be small values" [4, p44].

Suppose that the expression (29) obtained by the angular momentum method is valid. At this time, let us consider (37):

$$-\frac{\overline{\Delta T_{0E}(i^d)}}{T_{00E}} = m_z = \psi_z = -\frac{h_z}{\Omega_0 I} = -\frac{dK_\alpha}{\Omega_0 I} = -\frac{K_\alpha^{max}}{\Omega_0 I} \times \frac{dK_\alpha}{K_\alpha^{max}} \qquad (46)$$

In view of $m_z \cong \frac{dK_\alpha}{K_\alpha^{max}} \approx 4.4988 \times 10^{-8}$ in (37), we have $K_\alpha^{max} = I\Omega_0 = 5.8642 \times 10^{40}[gCm^2 s^{-1}]$.

It shows that the atmospheric angular momentum becomes equal to the large angular momentum of the body of the Earth, which leads to the physical contradiction.

If $\frac{dK_\alpha}{K_\alpha^{max}}$, the relative ratio of the atmospheric angular momentum becomes much less than $\sim 10^{-8}$ (it is realistic and natural in the sense of a natural phenomenon by itself), the authors can conclude that $K_\alpha^{max}$, the maximal value of the atmospheric angular momentum of the Earth will be greater than $I\Omega_0 = 5.8642 \times 10^{40}[gCm^2 s^{-1}]$ on the condition that $m_z$, an observation, is fixed in the range of $10^{-8} \sim 10^{-9}$ ($K_\alpha^{max} \gg I\Omega_0$). It is a logical contradiction because the atmospheric angular momentum cannot be greater than the angular momentum of the body of the Earth.

Therefore, the angular momentum method gives the contradictory results in physical meaning.

That is why, we think it is right to believe that the terms related to the atmospheric angular momentum in (17) should be varied in the lower range of $10^{-14}$ less than the maximum $10^{-13}$ as mentioned in (42).

From the analysis mentioned above, the authors can draw a conclusion that the atmosphere in (17) can not be the main factor of the excitation function $\psi_z$ and the term $\delta K_\alpha(i^d) = h_z$ representing the movement of the atmosphere is not the main factor which can cause the instabilities of the Earth rotation.

The viewpoint will be found later through the numerical values in [4],[5].

## IV. Effect of the variation in the Earth's moment of inertia ($\Delta I_{33}$) on the Earth's rotation

As mentioned above, Munk and Hassan studied the effect of $\Delta I_{33}$ within the length of day in 1961 [1].

They regarded the variation to be very small in comparison to the variation in the atmospheric angular momentum, so consideration of the term has not been taken into in (17) and no one had a doubt about it.

Rapid development of the observation instruments made it possible to find the cause of the instabilities of the Earth rotation. Discovery of the short period oscillations led scholars to a great interest in the variation in the moment of inertia.

You can take papers of Pilinik as typical examples [25],[26],[27]. He put an attention on the law of angular momentum conservation of the Earth rotation

$$C\omega = const \qquad (47)$$

from the assumption that the Earth is an isolated system.
Using the following expression

$$\frac{\Delta C}{C} = -\frac{\Delta \omega}{\omega} = \frac{\Delta t}{t} \qquad (48)$$

obtained from (38), he estimated the variation in the moment of inertia

$$\Delta C = C \times \frac{\Delta t}{t} \qquad (49)$$

[27], [28].

Hence, the authors will use some statements written on the papers of Pilinik for explanation. He stated in [28] that the oscillation with the period of $5.^d64$ can never be explained by the tidal effect. Also, he mentioned that the major reasons of the short period oscillation were not explained [25]-[29].

He said that the problem of the instabilities of the Earth rotation within one day and night was drawing a particular concern [26],[27] and the atmospheric friction on the Earth's surface could not explain the problem [26],[22],[12].

In addition, he said that the observations of the Earth's rotation rate within one day and night could be explained only by the irregular variation of the Earth's angular velocity of rotation [26].

So, as a way out, he focused on the variation in the moment of inertia of the Earth.

In addition, he said that $\Delta C$ is equal to $16 \times 10^{35} gCm^2$, $32 \times 10^{35} gCm^2$, $185 \times 10^{35} gCm^2$, $476 \times 10^{35} gCm^2$ for the waves with a period of $27.^d6$, $13.^d7$, $0.^d49862$, and $0.^d34102$ [27].

Also, he said that although such great variation, of course, in the polar moment of inertia within one day and night would make objections, no other reasons had been found for explanation of the instabilities of the Earth's rotation within one day and night yet [27].

In fact, in order to explain the *LOD* (the value of *D*) announced by the IERS in terms of the variations in the moment of inertia $\delta I_0'(i^d) = \overline{\Delta I_{33}}$ in (9), they should vary as much as $\sim 10^{37}\, gCm^2$ at each time.

Therefore, the author of [27] pointed out that whether such great variations would be really possible or not within one day and night would cause much doubt taking $C = 8.068 \times 10^{44}\, gCm^2$ into consideration [27], [32].

However, such great change in the polar moment of inertia is impossible actually to occur in the Earth.

Although the Earth undergoes elastic deformation and the Earth's moment of inertia changes continually due to all kinds of geophysical phenomena occurring inside and outside the Earth's body, it will have a secondary effect on the Earth rotation.

Consequently, the variations in the moment of inertia $\delta I_0'(i^d) = \overline{\Delta I_{33}}$ in (17) is eliminated from the list of the main factor.

### V. Effect of the Moment of External Force on the Earth Rotation

As mentioned above in a quantitative way, the atmospheric mechanical action, variation in the atmospheric angular momentum and the moment of inertia of the Earth do not have much effect on the Earth rotation.

Therefore, there exists $M_z$, only one term of external force which can influence the Earth rotation.

In order to solve the problem, it is convenient to use the equation of the Earth rotation.

It was shown that the atmosphere had power of about $2 \times 10^{15}\,W$ affecting the Earth rotation and the flow of ocean had power of about $10^{14}\,W$ [5].

Let us analyze how much the following quantity proposed in [4], [5]

$$\frac{dE}{dt} = C\omega \frac{d\omega}{dt} \approx 10^{14} \sim 10^{15}\,W \tag{50}$$

can have an effect on the instabilities of the Earth rotation.

It is convenient for us to calculate the physical quantities in *CGSE* system of units and transfer them to *MKSE* system of units if necessary. That is

$$\frac{dE}{dt} = C\omega\frac{d\omega}{dt} \approx (10^{21} \sim 10^{22})[erg/s] \tag{51}$$

Dividing two sides of the upper expression by $\omega$, you obtain the equation of Earth rotation in essence. That is,

$$\frac{1}{\omega}\frac{dE}{dt} = C\frac{d\omega}{dt} \approx \frac{1}{\omega}(10^{21} \sim 10^{22})erg/s = M_z \Rightarrow C\frac{d\omega}{dt} = M_z \tag{52}$$

, where $\omega$, the angular velocity of the Earth rotation is equal to $\frac{2\pi}{86400s}$, so the force moment $M_z$, produced by the atmospheric force is as follows.

$$C\frac{d\omega}{dt} = M_z = \frac{86400}{2\pi} \times (10^{21} \sim 10^{22})[erg] = 1.375 \times 10^{25\sim 26}[erg] = 1.375 \times 10^{18\sim 19}[J] \tag{53}$$

You can find the magnitude of the rotation moment in [30] in which the bulge torque magnitude reaches $\sim 1.5 \times 10^{18} Nm = \sim 1.5 \times 10^{18} J = \sim 1.5 \times 10^{25}[erg]$ for the lunar
tidal band and according to Bizouard and Lambert, the external torque is $\sim 10^{17} Nm = \sim 10^{17} J = \sim 10^{24}[erg]$.

The value of rotation moment is in the very low range in comparison to the moment of external force $\sim 2.63 \times 10^{33}[erg]$ found by the authors through (15).

That is,

$$C\frac{d\omega}{dt} = 1.375 \times 10^{25\sim 26}[erg] \tag{54}$$

$$C\frac{d\omega}{dt} = -2.63 \times 10^{33}[erg] \tag{55}$$

The force moment produced due to the effect of atmosphere is very small in comparison to $1.23 \times 10^{29}[erg]$, the force moment of the Moon on the Earth rotation.

The moment can be calculated from the solution of the Poisson equations [6].

In order to have the quantitative notion, let us see the other variation of the equation of the Earth rotation (15) giving the solution to (17)

$$\frac{d}{dt}(I_0'\omega_0) = M_Z^0 + \sum_j M_{j\omega}^{per} - \frac{dK_\alpha}{dt} \qquad (56)$$

$\sim 2.63\times10^{33} \quad \sim 2.63\times10^{33} \gg 1.23\times10^{29} \gg 1.375\times10^{25\sim26}\,[erg]$

In other words, the solution to the equation is (17).
For quantitative understanding, the numerical magnitude of terms was given below the terms in the expression.
Here $K_\alpha$ is the atmospheric absolute angular momentum, the magnitude of which can be expressed as shown in (32) as:

$$K_\alpha = \sum_i m_i R_i^2 \omega_{0i} \le K_\alpha^{max} = m_W R_E^2 \omega_z = I_W \omega_z = 1.5053\times 10^{35}\,[gCm^2 s^{-1}] \qquad (57)$$

$M_{j\omega}^{per}$ can be regarded as everything, you think, including the planets such as Moon affecting the Earth rotation. The largest term among them is the force moment caused by gravitational interaction of the Moon.

All of observations $m_z$ expressed in (17), in other words, all of observations listed in human history showing the instabilities of the Earth rotation can be explained in a quantitative way in terms of (56), the equation of the Earth's rotation.

Although the gravitational attraction of the Moon are ignored, the equation of the Earth's rotation (55) can explain the instabilities of the Earth rotation well.

Thus, even the effect of the Moon can be ignored in case of quantitative analysis of the instabilities of the Earth rotation.

Then how can mechanical action of the atmosphere cause them ?
The following expressions can give the answer.
For convenience of further calculation, at first let us calculate the angular acceleration of the Earth rotation that can be obtained by dividing both sides of (56) by $I_0'$, the moment of inertia of the Earth. That is,

$$\frac{d}{dt}(\omega_0) = \frac{1}{I_0'}M_Z^0 + \frac{1}{I_0'}\sum_j M_{j\omega}^{per} - \frac{1}{I_0'}\frac{dK_\alpha}{dt} \qquad (58)$$

The value of the expression is calculated as follows.

$$\frac{d\omega_0}{dt} = -\frac{3.26\times10^{33}}{8.068\times10^{44}} + \frac{1.23\times10^{29}}{8.068\times10^{44}} - \frac{1.375\times10^{25\sim26}}{8.068\times10^{44}}$$

$$\frac{d\omega_0}{dt} = -3.2598\times10^{-12} + 1.5245\times10^{-16} + \cdots -1.7042\times10^{-20\sim-19}$$

(59)

$$\frac{d\omega_0}{dt} = -3.2598\times10^{-12} \qquad (60)$$

The unit is ($rad/s^2$).

If we neglect the effect of the external force in (59), we can find the value of angular acceleration represented in [4], [5]

$$\frac{d\omega_0}{dt} = -1.7042\times[10^{-20} \sim 10^{-19}]\, rad/s^2 \qquad (61)$$

However, (59) shows that it cannot be so, because the numerical values mentioned above are much greater than $10^{-20} \sim 10^{-19}$.

Therefore, the mean of angular acceleration of the Earth rotation is described in (60) rather than in (61).

Next, let us take a look at the factor of the Earth rotation from the view of point of power (force) proposed in [4], [5].

To do this, multiply both sides of (56) by $\omega_0$.

The result can be written as:

$$\frac{dE}{dt} = \omega_0 \frac{d}{dt}(I_0'\omega_0) = \omega_0 M_Z^0 + \omega_0 \sum_j M_{j\omega}^{per} - \omega_0 \frac{dK_\alpha}{dt} \qquad (62)$$

The expression can be rewritten as:

$$\begin{array}{l}
\frac{dE}{dt} = \omega_0 \frac{d}{dt}(I_0'\omega_0) = \quad \omega_0 M_Z^0 \quad + \omega_0 \sum_j M_{j\omega}^{per} \quad - \omega_0 \frac{dK_\alpha}{dt} \\
CGSE \qquad\qquad\qquad 1.9126\times 10^{29},\quad 8.945\times 10^{24},\quad 10^{21} \sim 10^{22}\,[erg/s] \\
MKSE \qquad\qquad\qquad 1.9126\times 10^{22},\quad 8.945\times 10^{17},\quad 10^{14} \sim 10^{15}\,[W] \\
\qquad\qquad\qquad\text{Effect of the moment of force} \gg \text{effect of atmosphere, ocean and the Earth's magnetic field, etc.}
\end{array} \qquad (63)$$

For convenience, the corresponding numerical values were given below the terms of the expression.

As the previous researchers did, if you assume that the Earth is an isolated system, you can obtain the result from (63) reflecting the viewpoint (50) indicated in [4],[5]

$$\frac{dE}{dt} = \omega_0 \frac{d}{dt}(I_0'\omega_0) \approx \omega_0 C \frac{d\omega_0}{dt} \approx \omega_0 I_W \frac{d\omega_0}{dt} \approx 10^{14} \sim 10^{15}\, W \qquad (64)$$

The change in energy of rotation of the atmosphere can be calculated as follows.

$$\begin{array}{l}
\frac{d}{dt}(\frac{1}{2}I_W\omega_0^2) = \omega_0 I_W \frac{d\omega_0}{dt} = \omega_0 \frac{d(I_W\omega_0)}{dt} = \omega_0 \frac{dK_\alpha}{dt} \leq \\
\qquad \leq \frac{2\pi}{86400}\times 2.0701\times 10^{39} \times 3.2598\times 10^{-12} = \\
\qquad = 4.907\times 10^{23}\,[erg/s] = 4.907\times 10^{16}\,[W]
\end{array} \qquad (65)$$

Our calculation gives the value of maximal power that the rotary kinetic energy of the atmosphere can change, because the authors used the maximum of the moment of inertia ( $I_W = m_W \times R_E^2 = 2.07 \times 10^{39} [gCm^2]$ ) that the atmosphere can have.

However, in fact, the atmospheric angular momentum should be smaller than the value if it changes.

$10^{14} \sim 10^{15} W$ proposed in [4],[5] is one order smaller than the value $\sim 10^{16} W$ calculated by the authors.

Now that the effect of the Moon greater than the power (force) of $\sim 10^{16} W$ is ignored when taking the instabilities of the Earth rotation into consideration, how can the power (force) of $10^{14} \sim 10^{15} W$ proposed in [4], [5] have the real effect on the instabilities of the Earth rotation?

The authors can show that their effect was contained in the observed errors to measure the instabilities of the Earth rotation.

That is why the term related to the atmosphere in (62) as the factor affecting the instabilities of the Earth's rotation will be excluded.

All of the geophysical phenomena which have less effect than that of the atmosphere are excluded from the factors which can cause the instabilities of the Earth's rotation.

Thus, the geophysical phenomena including the atmosphere of the Earth cannot be considered to be the main factor causing the instabilities of the Earth rotation.

Then, what is the main factor causing them?

As you can see in (17) and (56), it is $M_z$, the moment of external force that the former investigators have ignored. That is, the major reason for the instabilities of the Earth rotation is the moment of external force **M** in the equation of the Earth rotation (1).

However, the previous investigators neglected it.

The fact shows clearly how it is wrong that they assumed the Earth was an isolated system.

Although the previous researchers put a great significance on the relative angular momentum of the atmosphere $h_z$ and wrapped it in silk, the former approaches led to the contradiction to the physical basis. So, they do not carry validity.

Of course, it is of importance that the former researchers found the relation of the atmospheric variation to the variation in the meteorological data and the instabilities of the Earth rotation [3], [4],[5],[30].

It is because $\frac{dK_\alpha}{dt}$, the variation in the atmospheric angular momentum is the geophysical phenomenon described as a function of the Earth's angular velocity $\omega_0$ from the point of view of the equation of the Earth's rotation (56) under the correlation with the physical quantities related to the Earth rotation as follows

$$-\frac{dK_\alpha}{dt} = \frac{d}{dt}(I_0^/ \omega_0) \quad - M_Z^0 \quad -\sum_j M_{j\omega}^{per} - \cdots = f[\omega_0, \cdots] \quad (66)$$
$$1.375 \times 10^{25\sim 26} \approx \sim 2.63 \times 10^{33} - 2.63 \times 10^{33}, \quad 1.23 \times 10^{29} \quad [erg]$$

In other words, as (13) and (66) show, the temporal variations of the atmosphere are very complicated and various physical phenomena which change unpredictably owing to the Earth's periodic revolution and rotation and gravitational interaction with other planets as well as the Earth's deformations, variations of temperature of the atmosphere, atmospheric pressure variations and so on.

Therefore, finding of the characteristics of the instabilities in the Earth's rotation through the observation of the atmospheric change of the Earth is very natural.

That is why irrespective of the relative atmospheric angular momentum the being obtained from the meteorological oscillational data or by some other ways, the explanation of the instabilities of the Earth rotation UT1 (LOD) by means of the relative atmospheric angular momentum is wrong.

Therefore, one should not study the cause of UT1 (LOD), the instabilities of the Earth rotation, based on the variation in the atmospheric angular momentum, whereas you should study and predict the variation in the atmospheric angular momentum based on the observation of and research in the instabilities of the Earth rotation, at the same time, should study all kinds of meteorological data caused by it. To do so will be the right direction to the research work in it.

### VI. The Law of the Earth Rotation Change and Its Prospect

As a matter of fact, the Earth is not an isolated system. The Earth is a planet which moves in the Sun's gravitational field and is in the gravitational interaction with the Moon and the other planets. Therefore, the main factor causing the instabilities of the Earth rotation is the moment by the external force as seen from the quantitative expressions, especially in (17) and (56).

The problem leads to the study in the Law of the Earth Rotation Change. An attempt was made to find the real law of the Earth rotation in [16].

However, the previous scientists have not found such Law of the Earth Rotation Change (14) as we have done [31],[32].

In fact, the history of study in theory of the Earth's rotation for 350 years has not recorded such the Law of the Earth Rota-tion Change(14) that we studied[33].

The solution of the Law of the Earth Rotation Change (14) makes it possible to explain not only instabilities of the Earth's rotation rate, but also precession and nutation of the Earth's rotation axis and polar motion in the body of the Earth in a quantitative way, while the causes for the astronomical phenomena can be proved by means of the factors in the Law of the Earth's Rotation Change in a quantitative way clearly.

In particular, the wonderful fact that although 17 astronomical constants entangled with each other, all the theoretical values following from the solution of the single equation were in good agreement with the astronomical observations shows clearly that we have discovered the real Law of the Earth's Rotation Change[31]. It was possible thanks to the valuable experience of the previous researchers. Therefore, the authors' findings lead to succession of the previous theory of the Earth's rotation.

The authors have developed a software package for scientific and technical service based on the Law of the Earth Rotation Change, which will help reduce the number of the various observation of the Earth rotation in large numbers and make a study in high level and in scientific and technological way, and thereby save funds needed for observation, study for the astronomic phenomena and application to them[32].

The authors hope that the Law of the Earth Rotation Change will be used for the mankind's happiness and welfare.

## VII. Conclusion

**Firstly,** the main factor that causes the instabilities of the Earth rotation is the moment by external forces acting on the Earth which gives a quantitative explanation of all the observations known at present as .described in (14).

**Secondly**, the instabilities of the Earth rotation are not due to mechanical action of the atmosphere, variation in the atmospheric angular momentum, all kinds of astronomic phenomena and variation in the moment of inertia of the body of the Earth.

**Thirdly**, it is the right way to the study in theory on the Earth's rotation and development to know well the Law of the Earth Rotation Change (14) discovered by applying relativistic dynamics to theory on the Earth's rotation and apply it successfully to the astronomical observation and practical problems, which can provide a hopeful future for investigation of the Earth's rotation.

## Acknowledgment

The valuable findings of research of the previous researchers who lavished all their efforts on the search for the main factor causing the instabilities of the Earth rotation led us to obtain the results.

Therefore, we would like to thank the previous researchers including Nikolay Sidorenkov who greatly helped us to draw the conclusion mentioned above.

## Disclosure of interest

The authors declare that they have no competing interests.